\documentstyle[twoside,fleqn,espcrc2,epsf]{article}

\newcommand{\third}{\mbox{\small $\frac{1}{3}$}}         
\newcommand{\msbar}{\mbox{\tiny $\overline{MS}$}}        
\newcommand{\mom}{\mbox{\tiny $MOM$}}                    
\newcommand{\rgi}{\mbox{\tiny $RGI$}}                    
\newcommand{\plaq}{\Box}                                 
\newcommand{\born}{\mbox{\tiny $BORN$}}                  

\def\lsim{\mathrel{\rlap{\lower4pt\hbox{\hskip1pt$\sim$}}
    \raise1pt\hbox{$<$}}}                
\def\gsim{\mathrel{\rlap{\lower4pt\hbox{\hskip1pt$\sim$}}
    \raise1pt\hbox{$>$}}}                

\title{
       \vspace{-3.65cm}                                     %
       {\normalsize DESY 01-169}     \\[-0.2cm]             
       {\normalsize November 2001}    \\                     
       \vspace{2.70cm}                                      
       Progress towards a lattice determination of
       (moments of) nucleon structure functions%
            \thanks{Talk given by R. Horsley at Lat01,
                    Berlin, Germany.}}                      

\author{S. Capitani%
           \address{John von Neumann Institute NIC / DESY Zeuthen,
                    D-15738 Zeuthen, Germany},
        M. G\"ockeler%
           \address{Institut f\"ur Theoretische Physik, Universit\"at
                    Regensburg, D-93040 Regensburg, Germany},
        R. Horsley$^{\rm a}$,
        D. Pleiter$^{\rm a}$,
        P. Rakow$^{\rm b}$,
        H. St\"uben%
           \address{Konrad-Zuse-Zentrum f\"ur Informationstechnik Berlin,
                    D-14195 Berlin, Germany}
        and
        G. Schierholz$^{\rm a,}$%
           \address{Deutsches Elektronen-Synchrotron DESY,
                    D-22603 Hamburg, Germany},
        \newline {\it QCDSF} Collaboration }
       
\begin{document}

\begin{abstract}
Using unimproved and non-perturbatively $O(a)$ improved Wilson fermions,
results are given for the three lowest moments of unpolarised
nucleon structure functions.
Renormalisation, chiral extrapolation and the continuum limit
of the matrix elements are briefly discussed. The simulations are performed
for both quenched and two flavours of unquenched fermions.
No obvious sign of deviation from linearity in the chiral extrapolations
are found. (This is most clearly seen in our quenched unimproved data,
which extends to lighter quark mass.)
Possible quenching effects also seem to be small.
The lowest moment thus remains too large, so
it seems to be necessary to reach smaller quark masses
in numerical simulations.
\end{abstract}

\maketitle

\setcounter{footnote}{0}


\section{INTRODUCTION}
\label{introduction}

Much of our knowledge about {\it QCD} and the structure of the nucleon has
been derived from Deep Inelastic Scattering ({\it DIS}\/) experiments,
either $lN \to lX$ ($l = e^-, \mu^-$) via the exchange of a photon or
$\nu_l n \to l X$, $\overline{\nu}_l p \to l X$ via $W^+$ or $W^-$
respectively. The cross section is determined by the structure functions
$F(x, Q^2) \equiv 2xF_1, F_2$ and additionally for neutrino {\it DIS},
$xF_3$. Considering non-singlet ({\it NS} or $F^p - F^n$) combinations only
to avoid any extra gluon terms, the operator product expansion
relates moments of these structure functions to nucleon matrix elements
$v_{n;NS}$ as
\begin{eqnarray}
   \int_0^1 dx x^{n-2} F^{NS}(x,Q^2)
      = E_{F;v_n;NS}^{\rgi}(Q) v_{n;NS}^{\rgi} ,
                                            \nonumber
\end{eqnarray}
where $E_{F;v_n;NS}^{\rgi}$ is the Wilson coefficient and
$v_{n;NS}^{\rgi}$ is proportional to the matrix element of
${\cal O}^{(u)}-{\cal O}^{(d)}$. ${\cal O}^{(q)}$ is a quark bilinear form,
involving a $\gamma$-matrix and $n-1$ covariant
derivatives (see \cite{gockeler95a} for our conventions).
This `renormalisation group invariant' ({\it RGI}\/) form
gives a clean separation between a non-perturbative {\it number}
$v_{n;NS}^{\rgi}$ (which can be computed on the lattice)
and a {\it function} $E_{F;v_n;NS}^{\rgi}(Q)$ (which
is perturbatively known). This is a possible {\it direct} comparison
between the experimental result and the lattice result.
More practical at present, however, is to use {\it parton density} functions
(e.g.\ {\it MRS}, \cite{martin95a}) determined from {\it global}
fits and related to the structure function by a {\it convolution}.
Finally, in a scheme ${\cal S}$ at scale $M$ then
$v_{n;NS}^{\rgi} \equiv \Delta Z^{\cal S}_{v_n}(M) v_{n;NS}^{\cal S}(M)$ with
\begin{eqnarray}
   \lefteqn{[\Delta Z_{\cal O}^{\cal S}(M)]^{-1} \equiv
     \left[ 2b_0 g^{\cal S}(M)^2 \right]^{- {d_{{\cal O};0}\over 2b_0}} } & &
                                            \nonumber  \\
        & & \times \exp{\left\{ \int_0^{g^{\cal S}(M)} d\xi
               \left[ {\gamma^{\cal S}_{\cal O}(\xi)
                          \over \beta^{\cal S}(\xi)} +
                      {d_{{\cal O};0}\over b_0 \xi} \right] \right\} } ,
                                            \nonumber
\end{eqnarray}
which in the $\overline{MS}$ scheme with $\Lambda^{\msbar}$ from
\cite{booth01a} at $M = 2\mbox{GeV}$
gives for quenched $0.732(9)$, $0.596(10)$, $0.534(13)$
($n = 2$, $3$, $4$ respectively)
while for unquenched, $[\Delta Z^{\msbar}_{v_2}]^{-1} \sim 0.695(10)$.


\section{LATTICE DETAILS}
\label{lattice_details}

The euclideanisation and form of the ${\cal O}^{(q)}$
has been described in \cite{gockeler95a}: $v_{2b}$
can be determined with nucleon momentum $\vec{p} = \vec{0}$, while $v_{2a}$,
$v_3$ and $v_4$ need a three-momentum with one non-zero component.
By considering the {\it NS} term the difficult to
compute one-quark-line-disconnected terms cancel.

A given operator can mix with three
different classes of operators (with the same quantum numbers
under the hypercubic group):
[A] of higher dimension;
[B] the same dimension; [C] lower dimension. At present the $O(a)$
improvement operators (class [A]) are only known for $v_2$. The associated
improvement coefficients are not completely known, \cite{capitani00a}.
For $v_3$ and $v_4$ there are additionally two mixing operators 
belonging to class [B] and for $v_4$ a further operator in class [C].

For $v_2$ we have found that the numerical values of the improvement
terms are much smaller than the operator itself; thus dropping them
will cause only an insignificant error.

Renormalisation can be considered in the $MOM$ scheme -- both perturbatively
and non-perturbatively, \cite{martinelli95a},
by defining $Z_{\cal O}^{\mom}$ from
\begin{eqnarray}
   \left[ 
    \langle q(p) | {\cal O}^{\mom} | q(p) \rangle \right] =
      \langle q(p) | {\cal O}^{\born} | q(p) \rangle ,
                                            \nonumber
\end{eqnarray}
at $M^2 = p^2$. Perturbation theory gives
\begin{eqnarray}
   \lefteqn{Z_{\cal O}^{\mom}(p,a) = 1 +} & &
                                     \nonumber \\
               & & g_0^2 \left[ d_{{\cal O};0} \ln(ap)
                           - B^{\mom}_{\cal O}(c_{sw})
                         \right] + O(g_0^4) ,
                                            \nonumber
\end{eqnarray}
where we have now computed $B^{\mom}_{\cal O}(c_{sw})$ for a general
value of $c_{sw}$, \cite{capitani00a,capitani01a}.
The perturbative renormalisation constant
for the first mixing term for $v_3$ has also been computed; it turns
out to be very small. Numerically this term is also smaller than
the $v_3$ matrix element so we shall also drop it.
Problems arise with the remaining mixing terms though: the Born term
(between quark states) vanishes, making their determination
using only quark states impossible.
At the present stage of development we shall simply ignore
these extra terms.

\begin{figure}[htb]
    \vspace*{-0.25in}
    \epsfxsize=7.00cm \epsfbox{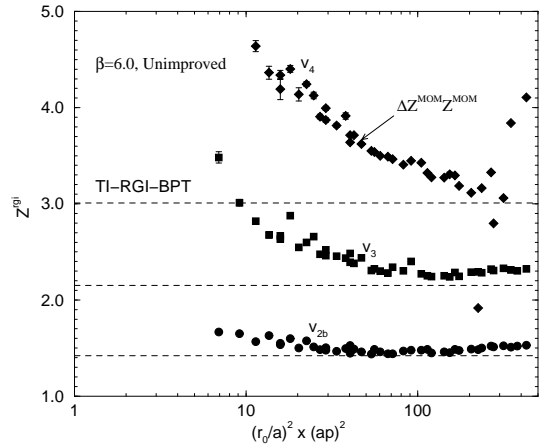}
    \vspace*{-0.30in}
    \caption{\footnotesize{\it $Z_{v_n}^{\rgi}$ using {\it TI-RGI-BPT}
             for unimproved quenched QCD (dashed lines)
             at $\beta = 6.0$.
             Also shown are the results from \cite{gockeler99a},
             filled symbols. (Expected is a window where the numbers
             are independent of $p^2$.)}}
   \vspace*{-0.30in}
   \label{fig_v2b_v3+v4_MOMms_nf0_rgi_lat01}
\end{figure}

Practically for the operator renormalisation we shall use a variant
of `tadpole improved, renormalisation group improved,
boosted perturbation theory', ({\it TI-RGI-BPT}),
\begin{eqnarray}
   {\cal O}^{\rgi} &\equiv& Z^{\rgi}_{\cal O} {\cal O}(a)
                                     \nonumber \\
       &\equiv& \Delta Z^{\mom}_{\cal O} Z^{\mom}_{\cal O} {\cal O}
        \equiv  \Delta Z^{\plaq}_{\cal O}(a) {\cal O}(a) ,
                                            \nonumber
\end{eqnarray}
\begin{eqnarray}
   \lefteqn{\Delta Z^{\plaq}_{\cal O}(a) = u_0^{1-n_D}
      \left[ 2 b_0 g^2_{\plaq} \right]^{d_{{\cal O};0}\over 2b_0} \times} & &
                                      \nonumber \\ 
     & & \left[ 1 + {b_1 \over b_0} g^2_{\plaq}
         \right]^{ { b_0 d_{{\cal O};1}^{\plaq} - b_1 d_{{\cal O};0}
                           \over 2 b_0b_1 }
                   + { {p_1 \over 4} {b_0 \over b_1} (1-n_D) } } ,
                                            \nonumber
\end{eqnarray}
(where $n_D$ is the number of derivatives in the operator,
$u_0^4 = \langle \third \mbox{Tr} U^{\plaq} \rangle$,
$g^2_{\plaq} = g_0^2 / u_0^4$ and $p_1$ is the first perturbative
coefficient in $u_0^4$). $d_{{\cal O};1}$ in the `$\plaq$' scheme
may be found from the known $d^{\mom}_{{\cal O};1}$ (from $O^{\rgi}$).
Results for unimproved quenched fermions are shown in
Fig.~\ref{fig_v2b_v3+v4_MOMms_nf0_rgi_lat01}
and compared with the non-perturbative method, \cite{gockeler99a}.
Reasonable agreement is seen.


\section{RESULTS}
\label{results}

\begin{figure}[htb]
    \epsfxsize=7.00cm \epsfbox{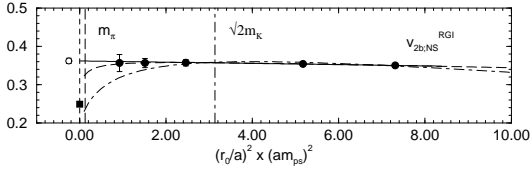}
    \vspace*{-0.30in}
    \caption{\footnotesize{\it $v^{\rgi}_{2b;NS}$ versus $(r_0 m_{ps})^2$
             for unimproved fermions at $\beta =6.0$.
             A linear fit gives 0.362(7). As further examples
             of fits we take the unquenched
             $d_n \to (3g_A^2 + 1) /(4\pi r_0 f_\pi)^2 \sim 0.663$.
             The dashed-dotted line is
             $\mu_\chi = 550\mbox{MeV}$ (fixed), the long-dashed
             line $\mu_\chi \sim 250 \mbox{MeV}$ (fitted).
             The filled square is from {\it MRS} \protect\cite{martin95a}.}}
   \vspace*{-0.25in}
   \label{fig_x1u-d_wilson_rgi_lat01}
\end{figure}

There have recently been suggestions
\cite{detmold01a,chen01a,arndt01a,chen01b}
for the behaviour of $v_n$ close to the chiral limit,
\begin{eqnarray}
   \begin{array}{l}
      v^{\rgi}_{n;NS} = C_n (r_0 m_{ps})^2 +  \\ \qquad
      B_n \left[ 1 - d_n (r_0 m_{ps})^2
                           \ln { (r_0 m_{ps})^2 \over
                                 (r_0 m_{ps})^2 + (r_0 \mu_\chi)^2 }
      \right]
   \end{array}
                                            \nonumber
\end{eqnarray}
rather than a linear form: $C_n (r_0 m_{ps})^2 + B_n$.
We check this using quenched unimproved results
(because this data extends down to lighter quark mass)
in Fig.~\ref{fig_x1u-d_wilson_rgi_lat01}.
Although not conclusive, it would seem that presently linear fits are adequate
and any possible non-linearities can only show up at rather small quark
mass outside the present range of data.

\begin{figure}[htb]
   \vspace*{-0.25in}
   \epsfxsize=7.00cm \epsfbox{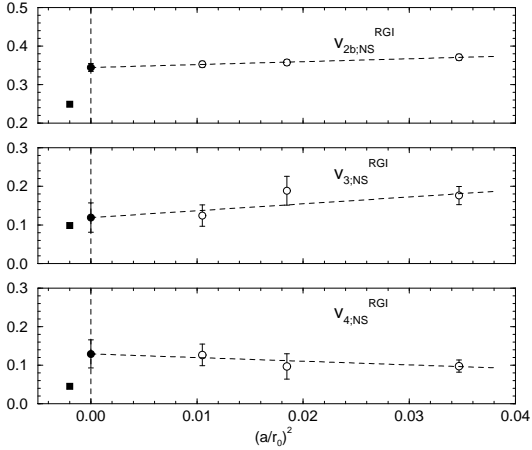}
   \vspace*{-0.30in}
   \caption{\footnotesize{\it Continuum extrapolation for
            quenched {\it QCD} for $v_{2b;NS}^{\rgi}$, $v_{3;NS}^{\rgi}$,
            and $v_{4;NS}^{\rgi}$, with
            continuum values 0.344(10), 0.119(38), 0.129(36) respectively.}}
   \vspace*{-0.25in}
   \label{fig_x1b+x2+x3_aor02_lat01}
\end{figure}

We first consider the continuum limit using quenched $O(a)$ improved
fermions (at $\beta =$ $6.0$, $6.2$, $6.4$) after taking the chiral limit.
In Fig.~\ref{fig_x1b+x2+x3_aor02_lat01}
we show the results. The data for the higher moments is
unfortunately rather noisy.

\begin{figure}[htb]
    \vspace*{-0.10in}
    \epsfxsize=7.00cm \epsfbox{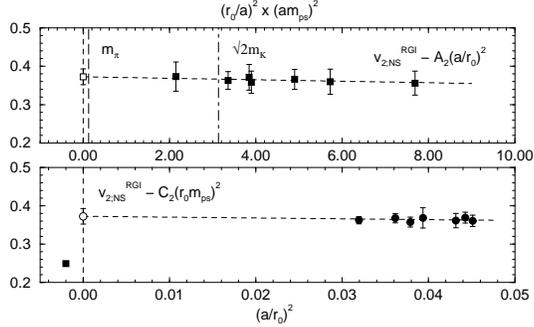}
    \vspace*{-0.30in}
    \caption{\footnotesize{\it Chiral extrapolation (upper picture)
             and continuum extrapolation (lower picture)
             for $v^{\rgi}_{2b;NS}$ using the data sets
             in \protect\cite{booth01a}, giving a result of 0.373(20).}}
   \vspace*{-0.25in}
   \label{fig_x1b_1u-1d.p0_mpi2+a2_nf2_lat01}
\end{figure}

Finally we consider {\it unquenched} results.
In Fig.~\ref{fig_x1b_1u-1d.p0_mpi2+a2_nf2_lat01} we use a fit function
ansatz
\begin{eqnarray}
   v_{2b;NS}^{\rgi} = A_2 \left( a / r_0 \right)^2 
                         + B_2 + C_2 \left( r_0 m_{ps} \right)^2 ,
                                            \nonumber
\end{eqnarray}
for $O(a)$ improved fermions with $7$ data sets. 
There is no obvious difference to the {\it quenched result} -- which
is surprising, \cite{dolgov00a}.
While for quenched {\it QCD} we expect that
the sea term is suppressed giving too large a quark contribution,
this is not so for unquenched {\it QCD}.

In conclusion the results presented here would seem to indicate
that we have to be much closer to the chiral limit in order
to be able to perceive the partonic properties
of the nucleon.
Further details will be given in \cite{capitani01a}.


\section*{ACKNOWLEDGEMENTS}

The numerical calculations were performed on the Hitachi {\it SR8000} at
LRZ (Munich), the Cray {\it T3E}s at EPCC (Edinburgh), NIC (J\"ulich) and
ZIB (Berlin) as well as the APE/Quadrics at NIC (Zeuthen).
We wish to thank the {\it UKQCD} Collaboration
for sharing their unquenched configurations with us.



\begin{thebibliography}{9}

\bibitem{gockeler95a}
   M. G\"ockeler et al.,
   Phys. Rev. D53 (1996) 2317, hep-lat/9508004.

\bibitem{martin95a}
   A.~D. Martin et al.,
   Phys. Lett. B354 (1995) 155, hep-ph/9502336.

\bibitem{booth01a}
   S. Booth et al., hep-lat/0103023.

\bibitem{capitani00a}
   S. Capitani et al.,
   Nucl. Phys. B593 (2001) 183, hep-lat/0007004.

\bibitem{capitani01a}
   S. Capitani et al., in preparation.

\bibitem{martinelli95a}
   G. Martinelli et al.,
   Nucl. Phys. B445 (1995) 81, hep-lat/9411010.

\bibitem{gockeler99a}
   M. G\"ockeler et al.,
   Nucl. Phys. B544 (1999) 699, hep-lat/9807044.

\bibitem{detmold01a}
   W. Detmold et al., hep-lat/0103006.

\bibitem{chen01a}
   J.-W. Chen et al., hep-ph/0105197.

\bibitem{arndt01a}
   D. Arndt et al., nucl-th/0105045.

\bibitem{chen01b}
   J.-W. Chen et al., nucl-th/0108042.

\bibitem{dolgov00a}
   D. Dolgov et al.,
   Nucl. Phys. Proc. Suppl. 94 (2001) 303,
   hep-lat/0011010.

\end{thebibliography}
\end{document}